\documentclass[aps,prl,twocolumn,showpacs,superscriptaddress,groupedaddress]{revtex4-1}  
\usepackage{graphicx}  
\usepackage{dcolumn}   
\usepackage{bm}        
\usepackage{amssymb}   
\usepackage{amsmath}
\usepackage{color}

\newcommand{\todo}[1]{\textcolor{black}{#1}}
\newcommand{\Tr}{\mathrm{Tr}}

\begin{document}

	\title{Geometry of wave propagation on active deformable surfaces}
	
	\author{Pearson W. Miller}
	\affiliation{Department of Mathematics, Massachusetts Institute of Technology, 77 Massachusetts Avenue, Cambridge,~MA~02139-4307, USA}
	
	\author{Norbert Stoop}
	\affiliation{Department of Mathematics, Massachusetts Institute of Technology, 77 Massachusetts Avenue, Cambridge,~MA~02139-4307, USA}

	\author{J\"orn Dunkel} 
	\affiliation{Department of Mathematics, Massachusetts Institute of Technology, 77 Massachusetts Avenue, Cambridge,~MA~02139-4307, USA}

	\date{\today}

\begin{abstract}
Fundamental biological and biomimetic processes, from tissue morphogenesis to soft robotics, rely on the propagation of chemical and mechanical surface waves to signal and coordinate active force generation. \todo{The complex interplay between surface geometry and contraction wave dynamics remains poorly understood, but will be essential for the future design of chemically-driven soft robots and active materials. Here, we couple prototypical chemical wave and reaction-diffusion models to non-Euclidean shell mechanics to identify and characterize generic features of chemo-mechanical wave propagation on active deformable surfaces.} Our theoretical framework is validated against recent data from contractile wave measurements on ascidian and starfish oocytes, producing good quantitative agreement in both cases. The theory is then applied to illustrate how geometry and preexisting discrete symmetries can be utilized to focus active elastic surface waves.  \todo{We highlight the practical potential of chemo-mechanical coupling by demonstrating spontaneous wave-induced locomotion of elastic shells of various geometries}. Altogether, our results show how geometry, elasticity and  chemical signaling can be harnessed to construct dynamically adaptable, \todo{autonomously moving} mechanical surface wave guides.  
\end{abstract}
	
	\maketitle

Wave propagation in complex geometries has been studied for centuries \cite{huygens1885traite} in fields as diverse as optics \cite{solli2007optical}, hydrodynamics \cite{bush2015pilot} or gravitation \cite{abbott2016observation}. The motion of a wave can be manipulated by precisely tuning the geometrical properties of its medium, an effect exploited by novel optical \cite{pendry2006controlling} and acoustic~\cite{wu2011elastic, wang2015topological} metamaterials with versatile refractive properties. \todo{The problem of wave guidance becomes particularly interesting in soft active systems, where travelling waves can locally contract, shear or otherwise deform the surfaces on which they propagate. Although most commonly seen in biological contexts \cite{beta2017intracellular, allard2013traveling, salbreux2017mechanics, oster1984mechanics}, wave-induced deformation is increasingly being explored in the context of soft materials engineering \cite{ionov2014hydrogel}, and there is evidence suggesting that such deformation can exert strong feedback on the propagation of chemical waves \cite{gov2006dynamics, PhysRevLett.103.238101, Frank2017diffusion}. Broadening our understanding of chemo-mechanical wave propagation is essential for the development of smart materials and soft robotics devices~\cite{hu2015buckling, wehner2016integrated} that utilize chemical gradients and targeted buckling~\cite{paulose2015selective}.}

\todo{Here, we show how the complex interplay between autonomous chemical wave guidance and geometry can be used to functionalize soft active matter. 
Our analysis  builds on a generic, broadly applicable model in which wave dynamics couples covariantly to a deformable elastic surface.  We first validate the theoretical framework by replicating the behavior of contraction waves observed in recent experiment~\cite{carroll2003exploring,Bischof_NCOMM,bischof2016molecular}. Subsequently, the theory is applied to design autonomously moving  elastic shells of various shapes.  Our results highlight that mechanical feedback can influence both the shape and speed of chemical waves, enabling significantly faster locomotion.} 

\begin{figure*}[t!]
\includegraphics[width=0.9\textwidth]{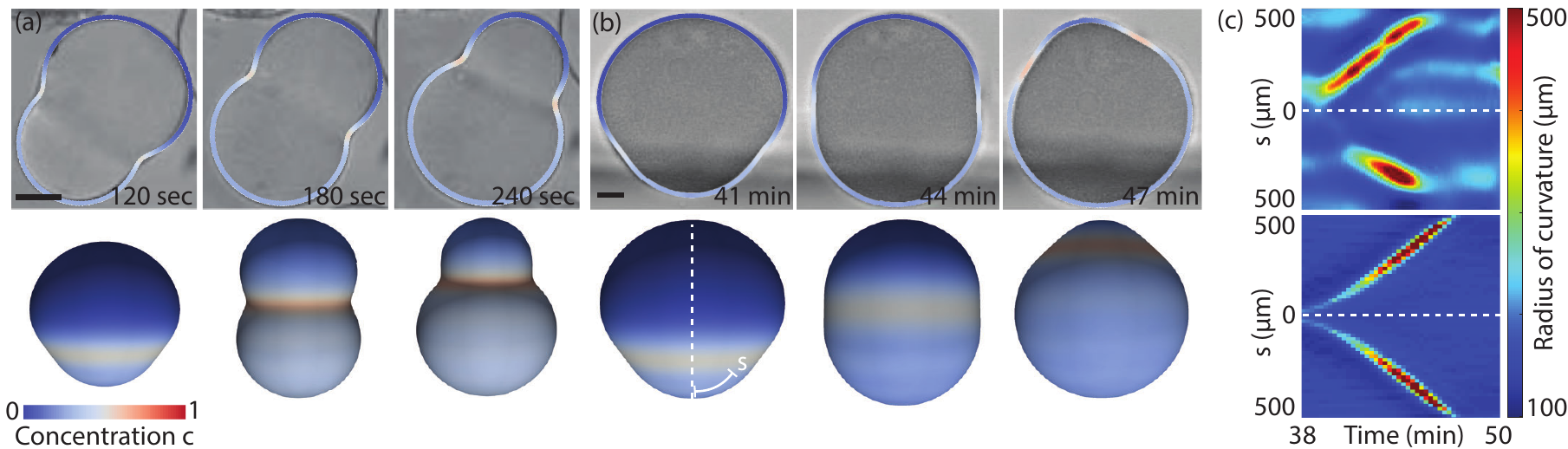}
\caption{\label{fig:fig1} Traveling wave model reproduces \textit{in vivo} mechanics of embryonic surface contraction waves.
(a)~For $A_{\textrm{C}} = 1.5$ and $A_{\textrm{I}} = 1.6$, the contraction wave model reproduces sperm-triggered ascidian embryo shape dynamics. Top: Microscopy images of the deformed oocytes (adapted with permission from Ref.~\cite{carroll2003exploring}), overlaid with cross-sections of our simulation. Bottom: Corresponding 3D surfaces indicating the concentration field. Scale bar 50 $\mu$m. See Movie 1 for full motion.   
(b)~Simulation with $A_{\textrm{C}} = A_{\textrm{I}} = 0.5$ reproduces  the surface contraction wave in starfish oocytes during anaphase in meiosis I. Experimental images adapted with permission from \cite{bischof2016molecular}. Scale bar 30 $\mu$m. See Movie 2 for full motion. (c)~Kymographs depicting the local radius of mean curvature for the experimental (top) and simulation (bottom) cases. Regions of maximal curvature indicate the center of the  wave and travel at a constant speed. $\mu_V=  24 {Y h }/{R^4}$, $h/R = 0.05$. } 
\end{figure*}						

\par
Our model contains two ingredients: An elastic shell described by the geometry and displacement of a two-dimensional (2D) surface $\omega$, and a scalar concentration field $c$ on $\omega$ equipped with wave-like dynamics. To capture the mechanics of the shell, we use the well-established Koiter  model~\cite{Ciarlet}, in which stresses within the shell are integrated along the thickness direction. Assuming a small thickness $h$, the shell's mechanical configuration is then entirely described by the geometry of its middle-surface~$\omega$. Equilibrium configurations correspond to minima of the elastic energy $\mathcal{E}_\mathrm{KS} =  \mathcal{E}_\mathrm{S} + \mathcal{E}_\mathrm{B} $ with stretching and bending contributions

\begin{subequations}\label{Energy_functional}
\begin{align}
\mathcal{E}_\mathrm{S} 
= \frac{Y h }{8 (1-\nu^2)} 
\int_{\bar{\omega}} d\omega & 
\bigl\{ (1-\nu) \Tr \bigl[\left( \textbf{a} - \bar{\textbf{a}} \right)^2\bigr] \label{eq:stretching}
 \\ 
 & \qquad
 + \nu \left[ \Tr \left(\textbf{a}-\bar{\textbf{a}}\right)\right]^2 \bigr\} \;, \nonumber\\
\mathcal{E}_\mathrm{B} 
= \frac{Y h^3 }{24 (1-\nu^2)} 
\int_{\bar{\omega}} d\omega & 
\bigl\{(1-\nu) \Tr \bigl[\left( \textbf{b} - \bar{\textbf{b}} \right)^2\bigr]  \label{eq:bending} \\
		& \qquad+ \nu \left[ \Tr  \left(\textbf{b}-\bar{\textbf{b}}\right) \right]^2  \bigr\}\;.  \nonumber	
\end{align}
\end{subequations}
Here, $\omega$ denotes the deformed shell geometry, characterized by the metric $\mathbf{a} = (a_{\alpha\beta})$ and curvature tensor $\mathbf{b} = (b_{\alpha\beta})$, with Greek indices henceforth running from $1$ to $2$. $d \omega$ is the surface area element, $Y$ the shell's Young's modulus, and $\nu$ the Poisson ratio ($\nu = 0.33$ throughout). $\bar{\textbf{a}}$ and $\bar{\mathbf{b}}$ are metric and curvature tensors of the reference shell geometry $\bar{\omega}$. The shell has minimal energy if its deformed surface~$\omega$ coincides with $\bar{\omega}$. Conventionally, $\bar{\omega}$ is identified with the undeformed, stress-free configuration of the shell. Active, stimulus-driven stresses can however be included in this framework by allowing local modifications of the reference configuration~\cite{pezzulla2017curvature, heer2017actomyosin}. The surface $\bar{\omega}$ then generally becomes non-Euclidean~\cite{klein2007shaping, Efrati2009}. Here, we consider a concentration-dependent modification of  metric and curvature tensor of the form
\begin{equation}\label{eq:AC}
\bar{\textbf{a}} \rightarrow \exp\left( - A_{\textrm{C}} c\right) \bar{\textbf{a}},\qquad
\bar{\textbf{b}} \rightarrow (1 - A_{\textrm{I}} c) \bar{\textbf{b}},
\end{equation}
where coefficients $ A_{\textrm{C}}$ and $ A_{\textrm{I}}$ have the units of inverse concentration. The exponential dependence of $\bar{\textbf{a}}$ ensures a positive definite reference metric for all values of $c$. Consequently, $1/A_{\textrm{C}}$ is the characteristic concentration scale associated with the decay of the reference metric to its minimal value of zero. Since no such constraint is necessary for the curvature tensor, we choose a linear coupling. \todo{We note that for a surface parametrized by $(\theta_1, \theta_2)$, the surface element area is explicitly given by $d\omega=\sqrt{\mathrm{det}(\mathbf{\bar{a}})}d (\theta_1, \theta_2)$. Thus, for $A_C c_0 \gg 1$ with $c_0$ a characteristic concentration scale such as the peak concentration, the stress-free state of a surface element will have an area approaching $0$. Conversely, when $A_C c_0 < 1$, only a small relative change of the surface area will be induced. Similarly, the curvature coupling will tend to produce reference curvature on the same order of the original surface when $|A_I c_0| < 1$, and will produce much higher curvature when $|A_I c_0| \gg 1$.}			
\par
It remains to define the dynamics of the concentration field $c$. In biological systems, chemical waves often feature highly idiosyncratic behavior and strong dependence on parameter choices \cite{beta2017intracellular}. Despite these specific and unique aspects, we expect that generic features hold in many such systems. \todo{A minimal model of chemical wave propagation is given} by the telegraphic equation 
	\begin{equation}\label{eq:telegraph}
		c_{tt} + \alpha c_t = \gamma^2 \nabla^2 c.
	\end{equation}
Equation~\eqref{eq:telegraph} combines wave-like and diffusive behavior~\cite{kac1974stochastic, thomson1854theory, masoliver1996finite}. Specifically, parameters $\alpha$ and $\gamma$ determine the degree of diffusivity and the wave speed, respectively.   Equation~\eqref{eq:telegraph} is implicitly coupled to the surface $\omega$ via the geometry-dependent Laplace-Beltrami operator $\nabla^2 c = \frac{1}{\sqrt{|a|} }\partial_\alpha \left(  a^{\alpha \beta} \sqrt{|a|} \partial_\beta  c \right)$. In the following, we choose $\gamma$ such that the time scale associated with the wave propagation, $\tau_w = L / \gamma$, is much greater than the elastic equilibration time scale $\tau_m = \sqrt{L^2 \rho / \mathcal{E}_\mathrm{KS}}$, with $\rho$ the material density.  Separating the time scales allows for the following numerical time-stepping strategy: We discretize the deformed surface $\omega$ as well as the concentration field $c$ by $\mathcal{C}^1$-continuous subdivision finite elements~\cite{Cirak}. Further, throughout we consider the weak dissipation regime $\alpha = 0.001 \ll \tau_w^{-1}$, to ensure that the wavefront remains coherent over observed timescales. 

\begin{figure*}[t!]
	\includegraphics[width= 0.9\textwidth]{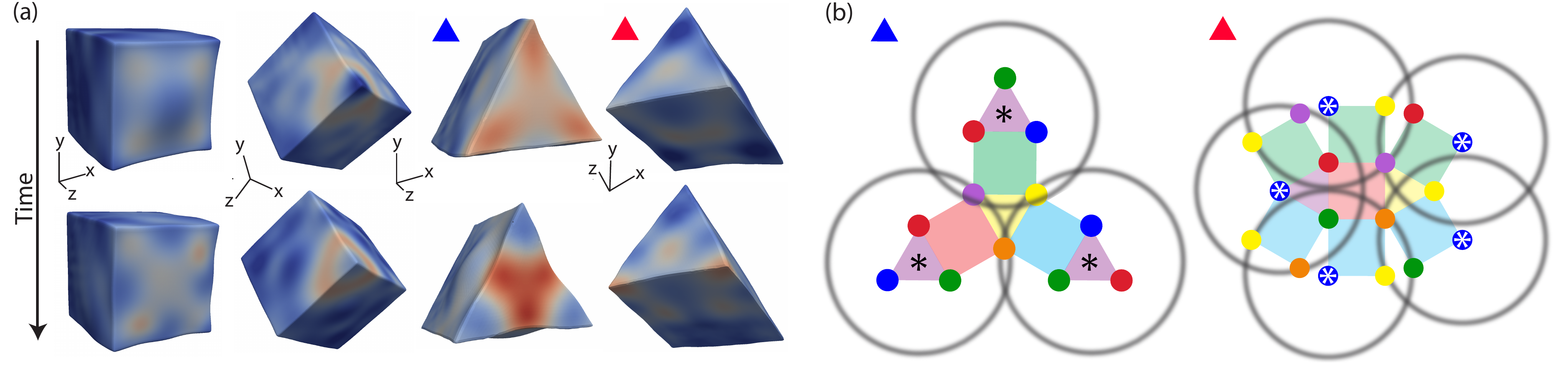}
	\caption{\label{fig:shape} For weak chemo-mechanical coupling, point wave propagation reflects the discrete symmetries of the elastic media. (a)~Weakly coupled ($A_{\textrm{I}} = 0.3$, $A_{\textrm{C}} = 1$, side-length $ L = 20 h$) waves propagating from a point on various geometries. On discrete surfaces, such as the cube and triangular prism shown here, self-interference increases concentration of the wavefront near edges and vertices and breaks the initial radial symmetry of the wave; see Movie 3. (b)~Unfolded representations of the two triangular prism cases shown in (a). Through choice of starting point $*$ , the wave front can be guided to exhibit a 3-fold or 5-fold symmetry on the opposing face.  Unique faces and vertices are color coded for clarity.}
\end{figure*}

\par
Solutions of Eq.~\eqref{eq:telegraph} do not necessarily conserve the total concentration. Integrating \eqref{eq:telegraph}  over an arbitrary, smooth simply-connected and closed surface gives $\ddot{\bar{c}} +\alpha \dot{\bar{c}}=0$ for the total concentration $\bar{c}(t)$. Thus, we can choose initial conditions in which the total concentration is uniformly increasing; this flexibility is essential for closely approximating experimentally measured chemical waves (Fig.~\ref{fig:fig1}). Unless stated otherwise, we use a narrow 2D Gaussian as our initial condition, given as $c(\phi, t = 0) = C_0 \exp[- \phi^2/(2 \sigma^2)]$ and $ \partial_t c(\phi, t = 0) = C_0 \phi \exp[-\phi^2/(2 \sigma^2)]$, where $\phi$ is the azimuthal angle, with the corresponding initial condition for the first derivative $\partial_t c(t = 0) = -\gamma \partial_\phi c(t = 0)$. These initial conditions were found to produce radially symmetric waves which maintained $c > 0$ everywhere as the wave expanded and then converged at the opposite pole (Movies~1, 2). Note that without loss of generality, we set the amplitude $C_0=1$, corresponding to a trivial rescaling of $c$ in Eq.~\eqref{eq:telegraph} and the coupling parameters $A_I, A_C$. Given an initial concentration field and shell geometry, we integrate Eq.~\eqref{eq:telegraph} using Verlet time-stepping and update the reference surface via Eq.~\eqref{eq:AC}. The shell configuration $\omega$ is then updated assuming overdamped dynamics, with forces calculated from the gradient of the shell energy, Eqs.~\eqref{eq:stretching} and~\eqref{eq:bending}~\cite{stoop2015curvature, heer2017actomyosin}. Due to the separation of wave and mechanical time scales, this approach ensures that the shell is very close to mechanical equilibrium at all times.

	The model defined by Eqs. \eqref{eq:bending}--\eqref{eq:AC} is generic and as such broadly applicable. \todo{To demonstrate its validity, we use it replicate one of the best-studied examples of contractile waves in nature, the single-cell contraction waves which occur during  oogenesis in many animal species \cite{cheer1987cortical}.} These occur shortly before the first cell division, when a contractile wave travels from the vegetal to the animal pole of the embryonic cell. Contraction is here driven by the localization and activation of myosin motors in the actin cortex. The dominant effect of the actin network contraction can be captured by a modification of the reference metric $\bar{\textbf{a}}$ alone \cite{taniguchi2013phase}. Important model systems for the study of this process are oocytes from the starfish \textit{Patiria miniata} and the ascidian \textit{Ascidiella aspersa} \cite{bischof2016molecular, bement2015activator, carroll2003exploring} (Fig.~\ref{fig:fig1}a,b, top row). \todo{Although actin cortex mechanics have studied in the physics community for decades, the above model is one of the first to incorporate a fully 3D model that incorporates active stress via a multiplicative framework, which is needed to properly treat large deformations \cite{goriely2008elastic}.} 
	
\par
Since the cell membrane is essentially impermeable over the wave propagation time scales, we assume that the enclosed fluid volume remains constant during the wave-induced contraction. To match our model with experiments, we thus augment Eq.~\eqref{Energy_functional} with an additional term $\mu_V (V - \bar{V})^2$ to penalize deviations of the internal volume $V$ from the reference volume~$\bar{V}$. Starting from a spherical oocyte shell and the initial concentration profile defined above, we tune our coupling parameters to produce local contraction and curvature of the magnitudes observed experimentally. The narrow Gaussian profile we use approximates the point-like initial conditions observed experimentally, with the thickness $\sigma$ chosen to match the wave thickness seen in experiment.  Overlaying cross sections from the elastic shell with microscopy images, we find excellent agreement between model and experiment (Fig.~\ref{fig:fig1}a,b). We note that the wave front remains rotationally symmetric during the entire process. To gain more insights into the dynamics, we construct kymographs depicting the spatial dynamics of curvature along cross section profiles (Fig.~\ref{fig:fig1}c). In both, simulations and experiments, the wave speed is roughly constant away from the poles (constant slopes in Fig.~\ref{fig:fig1}c), suggesting only a small influence of the metric contraction on the wave propagation dynamics. Near the origin of the wave, contraction is largely in-plane, effectively pulling the medium in the opposite direction of the wave propagation, while the wave maintains a constant velocity relative to local points on the surface, in the lab frame this distortion slows its progression. 

\begin{figure*}[t!]
\includegraphics[width=0.9\textwidth]{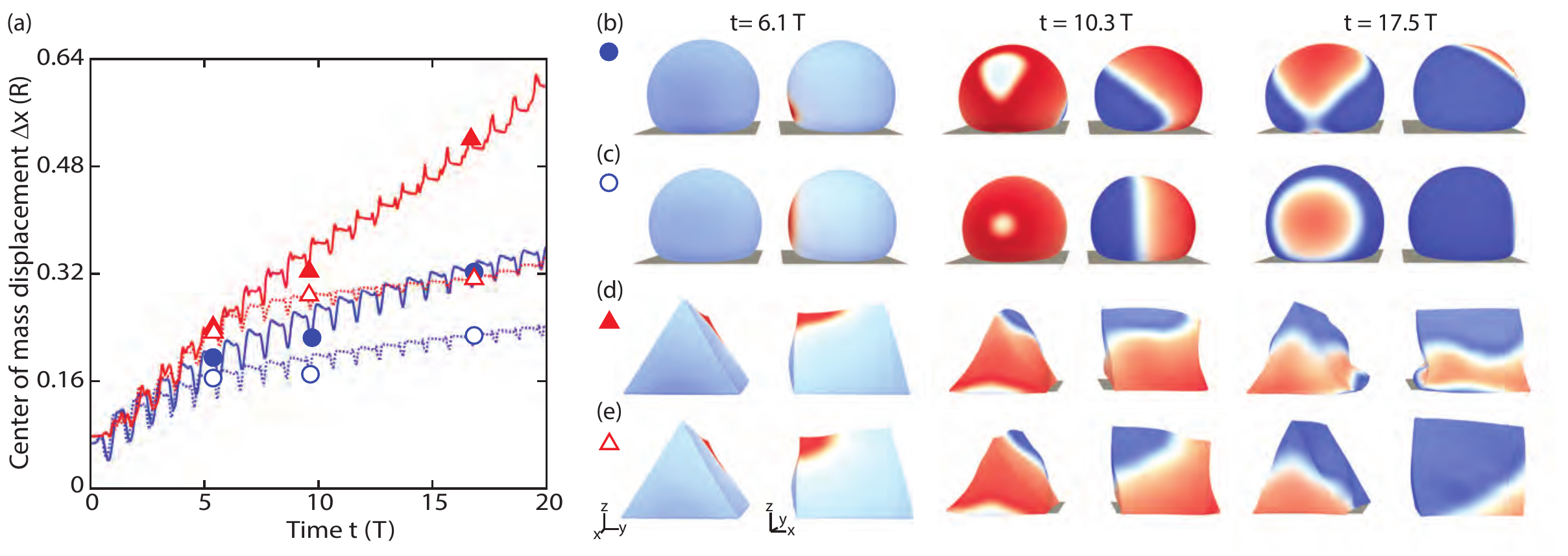}		
\todo{\caption{\label{fig:walking} 
Mechanically-coupled chemical waves induce locomotion in elastic shells on a solid substrate under gravity; see Movies~4-7. 
(a)~Locomotion in deformation-insensitive systems (hollow symbols, dashed lines) is slower than for deformation-coupled waves (full circles, solid lines) in spherical (blue) and prismatic shells (red). Time is normalized for each system in terms of oscillation period $T$, and distance is measured in terms of shell radius $R$. (b-e) Corresponding snapshots in time demonstrate that the wave shape is strongly affected by deformations 
(b,d)~as compared to deformation-insensitive wave propagation (c,e). Spherical shells have radius $R = 25$, while for prisms all edges have length $L = 2.3 R$. In all cases, thickness $h = 0.084 R$ and volume multiplier $\mu_V = 0.4 {Y h}/{R^4} = 11.1 {Y h}/{L^4}$.
}
 }
\end{figure*}	
	
Having validated the theoretical model on spherical shells, we study implications for wave propagation in more complex archetypal geometries. Choosing a regime where the deformations are relatively small compared to system size ($A_{\textrm{I}}=0.3, A_{\textrm{C}} = 1, \mu_V= 0$), we simulate contraction waves on cuboid and prismatic shells keeping  the same initial conditions as above,  but varying the location of wave origin. We find that waves start propagating radially outward but quickly disperse due to the broken rotational symmetry of the underlying geometry. Nonetheless, they converge again at later times, on the side of the surface opposite to the starting position. When they converge, they carry an imprint of the geometry of the underlying surface (Fig.~\ref{fig:shape}a). Since contraction is weak, this symmetry-breaking can be understood as follows: In the coupling-free limit $A_{\textrm{C}} \sim A_{\textrm{I}} \rightarrow 0$, equal-time points in the wave obey the Eikonal equation $\left| \nabla T \right| = \gamma^{-1}$, where $\gamma$ is the wave speed above, and $T(\bf r)$ is the time required for the wave front to reach geodesic distance $\bf r$ from the starting point at $t=0$  \cite{polthier1999geodesic}. 
For faceted surfaces, the Eikonal equation with point-like initial conditions can be solved graphically by unwrapping the faceted surface. For the prism surface and a wave emanating from the triangle face center (Fig.~\ref{fig:shape}a, column 3), we unfold the prism around the opposite triangular face (yellow face in Fig.~\ref{fig:shape}b, top) and periodically extend the unwrapped facets. Solutions of the Eikonal equation are then given as a superposition of circles centered in the origin of the wave (Fig.~\ref{fig:shape}b, top). Comparing with simulations, we find a qualitative agreement of the incident wave patterns also for waves originating from an edge on the prism (Fig.~\ref{fig:shape}a, column 4; Fig.~\ref{fig:shape}b, bottom), suggesting that basic geometry provides means to guide and shape surface waves. \todo{ Interestingly, we find waves on spherical shell with especially strong curvature inversion (where $A_I$ is sufficiently high that the modified preferred radius of curvature is smaller than $R$ by a factor of ten or more) produce a buckling pattern which induces discrete symmetry on the sphere. In spite of this change in global symmetry, we observe only small changes in the shape of the moving wave  (SI~\cite{SI}), demonstrating that global geometry alone is a poor predictor of wave dynamics on an active surface. 
}
\par
\todo{
To understand better  how sustained wave dynamics interact with geometry, deformation, and wave propagation, we consider an elastic shell resting on a flat surface under gravity (Fig.~\ref{fig:walking}). We model sustained, pulse-like wave dynamics by generalizing from the telegraph equation~\eqref{eq:telegraph} to the classic FitzHugh-Nagumo type model \cite{jones1984stability}
\begin{subequations} \label{FH-N}
	\begin{align}
		c_t &= \gamma^2 \nabla^2 c +  c (1 - c)(c - \nu)  - w \\
		w_t &= \epsilon ( c - \kappa w)
	\end{align}
\end{subequations} 
choosing parameter values within the regime of stable traveling wave solutions~\cite{jones1984stability}: $\nu = -0.01$, $\kappa = 0.0001$, $\epsilon = 0.005$,  and $\gamma = 0.2$. We couple $c$ to the local metric via Eq. \eqref{eq:AC} and fix a narrow Gaussian profile as initial condition for $c_0$, while $w(0) = 0$. Gravity is modeled by a constant field $\bf{F}_g = \rho g \hat{z}$, with $g = 0.5$ in our units, and $\rho = 0.01$ is the shell surface mass density. Contact forces with the ground are described in the SI \cite{SI}. 	
}
\par
\todo{
Our simulations show that each pulse leads to a small net displacement of the entire shell in the direction of wave propagation, resulting in a persistent motion over succeeding wave cycles (Fig.~\ref{fig:walking}a). Remarkably, even deformations that are small relative to the system size affect the wave dynamics considerably. To demonstrate this effect, we  consider a wave moving on a deforming sphere with $A_C = 0.1, A_I = 0$ (Fig.~\ref{fig:walking}b), and compare with a hypothetical reference scenario in which the wave deforms the sphere but keeps propagating on the undeformed reference surface geometry (Fig.~\ref{fig:walking}c, Movie 4). As can be seen in Fig.~\ref{fig:walking}a, the deformation-insensitive reference  sphere (blue dotted line) moves at a significantly reduced speed compared to the deformation-sensitive system (blue solid bline). Thus, mechanical feedback both increases the wave frequency by $\sim 10 \%$ and increases the speed of locomotion (Fig.~\ref{fig:walking}a).
}
\par
\todo{
The difference between deformation-sensitive and deformation-insensitive waves becomes even more striking in our second example, where we initiate the wave with $A_C = 0.2, A_I = 0$ from the vertex of a triangular prism (Fig.~\ref{fig:walking}d,e). Specifically, in the deformation-insensitive case, we obtain a net \lq walking\rq{} speed of less than half that of the deformation-sensitive system (Fig.~\ref{fig:walking}a). Moreover, we find that the shape and propagation mode of the wavefront is considerably altered due to deformation (Fig.~\ref{fig:walking}d,e; Movie 5). Additional movies showing sustained waves for other geometries and initial conditions are included in the SI~\cite{SI}. 
}	
\par
\todo{
To conclude, we have shown that a generic minimal model coupling dispersive chemical wave propagation with surface elasticity can reproduce quantitatively the experimentally observed surface deformation in ascidian and starfish oocytes (Fig.~1; Movies 1, 2). It was further demonstrated that waves confined to 3D embedded surfaces are highly susceptible to the underlying geometry of their surface (Fig.~2; Movie 3), results which could be experimentally confirmed by oocytes or reconstituted actin cortices confined in PDMS shaped cavities \cite{shah2014symmetry, bischof2016molecular,2016arXiv160307600T}. Finally, we showed that mechanical feedback caused by wave-induced deformation likewise has a clear effect on surface wave dynamics. The presented results on wave-induced shell locomotion should provide relevant insights for the future design of autonomous BZ-driven hydrogel actuators, which have already been shown capable of locomotion under externally prescribed contraction dynamics~\cite{yoshida2010self}. 
}

\par
This work was supported by a National Defense Science and Engineering Graduate Fellowship (P.M.), an Alfred P. Sloan
Research Fellowship (J.D.), an Edmund F. Kelly Research Award (J.D.), and a Complex
Systems Scholar Award of the James S. McDonnell Foundation (J.D.).

\appendix


\begin{thebibliography}{44}%
\makeatletter
\providecommand \@ifxundefined [1]{%
 \@ifx{#1\undefined}
}%
\providecommand \@ifnum [1]{%
 \ifnum #1\expandafter \@firstoftwo
 \else \expandafter \@secondoftwo
 \fi
}%
\providecommand \@ifx [1]{%
 \ifx #1\expandafter \@firstoftwo
 \else \expandafter \@secondoftwo
 \fi
}%
\providecommand \natexlab [1]{#1}%
\providecommand \enquote  [1]{``#1''}%
\providecommand \bibnamefont  [1]{#1}%
\providecommand \bibfnamefont [1]{#1}%
\providecommand \citenamefont [1]{#1}%
\providecommand \href@noop [0]{\@secondoftwo}%
\providecommand \href [0]{\begingroup \@sanitize@url \@href}%
\providecommand \@href[1]{\@@startlink{#1}\@@href}%
\providecommand \@@href[1]{\endgroup#1\@@endlink}%
\providecommand \@sanitize@url [0]{\catcode `\\12\catcode `\$12\catcode
  `\&12\catcode `\#12\catcode `\^12\catcode `\_12\catcode `\%12\relax}%
\providecommand \@@startlink[1]{}%
\providecommand \@@endlink[0]{}%
\providecommand \url  [0]{\begingroup\@sanitize@url \@url }%
\providecommand \@url [1]{\endgroup\@href {#1}{\urlprefix }}%
\providecommand \urlprefix  [0]{URL }%
\providecommand \Eprint [0]{\href }%
\providecommand \doibase [0]{http://dx.doi.org/}%
\providecommand \selectlanguage [0]{\@gobble}%
\providecommand \bibinfo  [0]{\@secondoftwo}%
\providecommand \bibfield  [0]{\@secondoftwo}%
\providecommand \translation [1]{[#1]}%
\providecommand \BibitemOpen [0]{}%
\providecommand \bibitemStop [0]{}%
\providecommand \bibitemNoStop [0]{.\EOS\space}%
\providecommand \EOS [0]{\spacefactor3000\relax}%
\providecommand \BibitemShut  [1]{\csname bibitem#1\endcsname}%
\let\auto@bib@innerbib\@empty
\bibitem [{\citenamefont {Huygens}(1885)}]{huygens1885traite}%
  \BibitemOpen
  \bibfield  {author} {\bibinfo {author} {\bibfnamefont {C.}~\bibnamefont
  {Huygens}},\ }\href@noop {} {\emph {\bibinfo {title} {Trait{\'e} de la
  lumi{\`e}re}}}\ (\bibinfo  {publisher} {chez Pierre Van der Aa, marchand
  libraire},\ \bibinfo {year} {1885})\BibitemShut {NoStop}%
\bibitem [{\citenamefont {Solli}\ \emph {et~al.}(2007)\citenamefont {Solli},
  \citenamefont {Ropers}, \citenamefont {Koonath},\ and\ \citenamefont
  {Jalali}}]{solli2007optical}%
  \BibitemOpen
  \bibfield  {author} {\bibinfo {author} {\bibfnamefont {D.}~\bibnamefont
  {Solli}}, \bibinfo {author} {\bibfnamefont {C.}~\bibnamefont {Ropers}},
  \bibinfo {author} {\bibfnamefont {P.}~\bibnamefont {Koonath}}, \ and\
  \bibinfo {author} {\bibfnamefont {B.}~\bibnamefont {Jalali}},\ }\href@noop {}
  {\bibfield  {journal} {\bibinfo  {journal} {Nature}\ }\textbf {\bibinfo
  {volume} {450}},\ \bibinfo {pages} {1054} (\bibinfo {year}
  {2007})}\BibitemShut {NoStop}%
\bibitem [{\citenamefont {Bush}(2015)}]{bush2015pilot}%
  \BibitemOpen
  \bibfield  {author} {\bibinfo {author} {\bibfnamefont {J.~W.~M.}\
  \bibnamefont {Bush}},\ }\href@noop {} {\bibfield  {journal} {\bibinfo
  {journal} {Annu. Rev. Fluid Mech.}\ }\textbf {\bibinfo {volume} {47}},\
  \bibinfo {pages} {269} (\bibinfo {year} {2015})}\BibitemShut {NoStop}%
\bibitem [{\citenamefont {Abbott}\ \emph {et~al.}(2016)\citenamefont {Abbott},
  \citenamefont {Abbott}, \citenamefont {Abbott}, \citenamefont {Abernathy},
  \citenamefont {Acernese}, \citenamefont {Ackley}, \citenamefont {Adams},
  \citenamefont {Adams}, \citenamefont {Addesso}, \citenamefont {Adhikari}
  \emph {et~al.}}]{abbott2016observation}%
  \BibitemOpen
  \bibfield  {author} {\bibinfo {author} {\bibfnamefont {B.~P.}\ \bibnamefont
  {Abbott}}, \bibinfo {author} {\bibfnamefont {R.}~\bibnamefont {Abbott}},
  \bibinfo {author} {\bibfnamefont {T.}~\bibnamefont {Abbott}}, \bibinfo
  {author} {\bibfnamefont {M.}~\bibnamefont {Abernathy}}, \bibinfo {author}
  {\bibfnamefont {F.}~\bibnamefont {Acernese}}, \bibinfo {author}
  {\bibfnamefont {K.}~\bibnamefont {Ackley}}, \bibinfo {author} {\bibfnamefont
  {C.}~\bibnamefont {Adams}}, \bibinfo {author} {\bibfnamefont
  {T.}~\bibnamefont {Adams}}, \bibinfo {author} {\bibfnamefont
  {P.}~\bibnamefont {Addesso}}, \bibinfo {author} {\bibfnamefont
  {R.}~\bibnamefont {Adhikari}},  \emph {et~al.},\ }\href@noop {} {\bibfield
  {journal} {\bibinfo  {journal} {Phys. Rev. Lett.}\ }\textbf {\bibinfo
  {volume} {116}},\ \bibinfo {pages} {061102} (\bibinfo {year}
  {2016})}\BibitemShut {NoStop}%
\bibitem [{\citenamefont {Pendry}\ \emph {et~al.}(2006)\citenamefont {Pendry},
  \citenamefont {Schurig},\ and\ \citenamefont
  {Smith}}]{pendry2006controlling}%
  \BibitemOpen
  \bibfield  {author} {\bibinfo {author} {\bibfnamefont {J.~B.}\ \bibnamefont
  {Pendry}}, \bibinfo {author} {\bibfnamefont {D.}~\bibnamefont {Schurig}}, \
  and\ \bibinfo {author} {\bibfnamefont {D.~R.}\ \bibnamefont {Smith}},\
  }\href@noop {} {\bibfield  {journal} {\bibinfo  {journal} {Science}\ }\textbf
  {\bibinfo {volume} {312}},\ \bibinfo {pages} {1780} (\bibinfo {year}
  {2006})}\BibitemShut {NoStop}%
\bibitem [{\citenamefont {Wu}\ \emph {et~al.}(2011)\citenamefont {Wu},
  \citenamefont {Lai},\ and\ \citenamefont {Zhang}}]{wu2011elastic}%
  \BibitemOpen
  \bibfield  {author} {\bibinfo {author} {\bibfnamefont {Y.}~\bibnamefont
  {Wu}}, \bibinfo {author} {\bibfnamefont {Y.}~\bibnamefont {Lai}}, \ and\
  \bibinfo {author} {\bibfnamefont {Z.-Q.}\ \bibnamefont {Zhang}},\ }\href@noop
  {} {\bibfield  {journal} {\bibinfo  {journal} {Phys. Rev. Lett.}\ }\textbf
  {\bibinfo {volume} {107}},\ \bibinfo {pages} {105506} (\bibinfo {year}
  {2011})}\BibitemShut {NoStop}%
\bibitem [{\citenamefont {Wang}\ \emph {et~al.}(2015)\citenamefont {Wang},
  \citenamefont {Lu},\ and\ \citenamefont {Bertoldi}}]{wang2015topological}%
  \BibitemOpen
  \bibfield  {author} {\bibinfo {author} {\bibfnamefont {P.}~\bibnamefont
  {Wang}}, \bibinfo {author} {\bibfnamefont {L.}~\bibnamefont {Lu}}, \ and\
  \bibinfo {author} {\bibfnamefont {K.}~\bibnamefont {Bertoldi}},\ }\href@noop
  {} {\bibfield  {journal} {\bibinfo  {journal} {Phys. Rev. Lett.}\ }\textbf
  {\bibinfo {volume} {115}},\ \bibinfo {pages} {104302} (\bibinfo {year}
  {2015})}\BibitemShut {NoStop}%
\bibitem [{\citenamefont {Beta}\ and\ \citenamefont
  {Kruse}(2017)}]{beta2017intracellular}%
  \BibitemOpen
  \bibfield  {author} {\bibinfo {author} {\bibfnamefont {C.}~\bibnamefont
  {Beta}}\ and\ \bibinfo {author} {\bibfnamefont {K.}~\bibnamefont {Kruse}},\
  }\href@noop {} {\bibfield  {journal} {\bibinfo  {journal} {Annu. Rev.
  Condens. Matter Phys.}\ }\textbf {\bibinfo {volume} {8}},\ \bibinfo {pages}
  {239} (\bibinfo {year} {2017})}\BibitemShut {NoStop}%
\bibitem [{\citenamefont {Allard}\ and\ \citenamefont
  {Mogilner}(2013)}]{allard2013traveling}%
  \BibitemOpen
  \bibfield  {author} {\bibinfo {author} {\bibfnamefont {J.}~\bibnamefont
  {Allard}}\ and\ \bibinfo {author} {\bibfnamefont {A.}~\bibnamefont
  {Mogilner}},\ }\href@noop {} {\bibfield  {journal} {\bibinfo  {journal}
  {Curr. Opin. Cell Biol.}\ }\textbf {\bibinfo {volume} {25}},\ \bibinfo
  {pages} {107} (\bibinfo {year} {2013})}\BibitemShut {NoStop}%
\bibitem [{\citenamefont {Salbreux}\ and\ \citenamefont
  {J{\"u}licher}(2017)}]{salbreux2017mechanics}%
  \BibitemOpen
  \bibfield  {author} {\bibinfo {author} {\bibfnamefont {G.}~\bibnamefont
  {Salbreux}}\ and\ \bibinfo {author} {\bibfnamefont {F.}~\bibnamefont
  {J{\"u}licher}},\ }\href@noop {} {\bibfield  {journal} {\bibinfo  {journal}
  {Phys. Rev. E}\ }\textbf {\bibinfo {volume} {96}},\ \bibinfo {pages} {032404}
  (\bibinfo {year} {2017})}\BibitemShut {NoStop}%
\bibitem [{\citenamefont {Oster}\ and\ \citenamefont
  {Odell}(1984)}]{oster1984mechanics}%
  \BibitemOpen
  \bibfield  {author} {\bibinfo {author} {\bibfnamefont {G.~F.}\ \bibnamefont
  {Oster}}\ and\ \bibinfo {author} {\bibfnamefont {G.}~\bibnamefont {Odell}},\
  }\href@noop {} {\bibfield  {journal} {\bibinfo  {journal} {Cytoskeleton}\
  }\textbf {\bibinfo {volume} {4}},\ \bibinfo {pages} {469} (\bibinfo {year}
  {1984})}\BibitemShut {NoStop}%
\bibitem [{\citenamefont {Ionov}(2014)}]{ionov2014hydrogel}%
  \BibitemOpen
  \bibfield  {author} {\bibinfo {author} {\bibfnamefont {L.}~\bibnamefont
  {Ionov}},\ }\href@noop {} {\bibfield  {journal} {\bibinfo  {journal} {Mater.
  Today}\ }\textbf {\bibinfo {volume} {17}},\ \bibinfo {pages} {494} (\bibinfo
  {year} {2014})}\BibitemShut {NoStop}%
\bibitem [{\citenamefont {Gov}\ and\ \citenamefont
  {Gopinathan}(2006)}]{gov2006dynamics}%
  \BibitemOpen
  \bibfield  {author} {\bibinfo {author} {\bibfnamefont {N.~S.}\ \bibnamefont
  {Gov}}\ and\ \bibinfo {author} {\bibfnamefont {A.}~\bibnamefont
  {Gopinathan}},\ }\href@noop {} {\bibfield  {journal} {\bibinfo  {journal}
  {Biophys. J.}\ }\textbf {\bibinfo {volume} {90}},\ \bibinfo {pages} {454}
  (\bibinfo {year} {2006})}\BibitemShut {NoStop}%
\bibitem [{\citenamefont {Chen}\ \emph {et~al.}(2009)\citenamefont {Chen},
  \citenamefont {Tsai}, \citenamefont {Wang},\ and\ \citenamefont
  {Lee}}]{PhysRevLett.103.238101}%
  \BibitemOpen
  \bibfield  {author} {\bibinfo {author} {\bibfnamefont {C.-H.}\ \bibnamefont
  {Chen}}, \bibinfo {author} {\bibfnamefont {F.-C.}\ \bibnamefont {Tsai}},
  \bibinfo {author} {\bibfnamefont {C.-C.}\ \bibnamefont {Wang}}, \ and\
  \bibinfo {author} {\bibfnamefont {C.-H.}\ \bibnamefont {Lee}},\ }\href
  {\doibase 10.1103/PhysRevLett.103.238101} {\bibfield  {journal} {\bibinfo
  {journal} {Phys. Rev. Lett.}\ }\textbf {\bibinfo {volume} {103}},\ \bibinfo
  {pages} {238101} (\bibinfo {year} {2009})}\BibitemShut {NoStop}%
\bibitem [{\citenamefont {Frank}\ \emph {et~al.}()\citenamefont {Frank},
  \citenamefont {Guven}, \citenamefont {Kardar},\ and\ \citenamefont
  {Shackleton}}]{Frank2017diffusion}%
  \BibitemOpen
  \bibfield  {author} {\bibinfo {author} {\bibfnamefont {J.~R.}\ \bibnamefont
  {Frank}}, \bibinfo {author} {\bibfnamefont {J.}~\bibnamefont {Guven}},
  \bibinfo {author} {\bibfnamefont {M.}~\bibnamefont {Kardar}}, \ and\ \bibinfo
  {author} {\bibfnamefont {H.}~\bibnamefont {Shackleton}},\ }\href@noop {}
  {\bibinfo  {journal} {arXiv:1710.00103}\ }\BibitemShut {NoStop}%
\bibitem [{\citenamefont {Hu}\ and\ \citenamefont
  {Burgue{\~n}o}(2015)}]{hu2015buckling}%
  \BibitemOpen
\bibfield  {journal} {  }\bibfield  {author} {\bibinfo {author} {\bibfnamefont
  {N.}~\bibnamefont {Hu}}\ and\ \bibinfo {author} {\bibfnamefont
  {R.}~\bibnamefont {Burgue{\~n}o}},\ }\href@noop {} {\bibfield  {journal}
  {\bibinfo  {journal} {Smart Mater. Struct.}\ }\textbf {\bibinfo {volume}
  {24}},\ \bibinfo {pages} {063001} (\bibinfo {year} {2015})}\BibitemShut
  {NoStop}%
\bibitem [{\citenamefont {Wehner}\ \emph {et~al.}(2016)\citenamefont {Wehner},
  \citenamefont {Truby}, \citenamefont {Fitzgerald}, \citenamefont {Mosadegh},
  \citenamefont {Whitesides}, \citenamefont {Lewis},\ and\ \citenamefont
  {Wood}}]{wehner2016integrated}%
  \BibitemOpen
  \bibfield  {author} {\bibinfo {author} {\bibfnamefont {M.}~\bibnamefont
  {Wehner}}, \bibinfo {author} {\bibfnamefont {R.~L.}\ \bibnamefont {Truby}},
  \bibinfo {author} {\bibfnamefont {D.~J.}\ \bibnamefont {Fitzgerald}},
  \bibinfo {author} {\bibfnamefont {B.}~\bibnamefont {Mosadegh}}, \bibinfo
  {author} {\bibfnamefont {G.~M.}\ \bibnamefont {Whitesides}}, \bibinfo
  {author} {\bibfnamefont {J.~A.}\ \bibnamefont {Lewis}}, \ and\ \bibinfo
  {author} {\bibfnamefont {R.~J.}\ \bibnamefont {Wood}},\ }\href@noop {}
  {\bibfield  {journal} {\bibinfo  {journal} {Nature}\ }\textbf {\bibinfo
  {volume} {536}},\ \bibinfo {pages} {451} (\bibinfo {year}
  {2016})}\BibitemShut {NoStop}%
\bibitem [{\citenamefont {Paulose}\ \emph {et~al.}(2015)\citenamefont
  {Paulose}, \citenamefont {Meeussen},\ and\ \citenamefont
  {Vitelli}}]{paulose2015selective}%
  \BibitemOpen
  \bibfield  {author} {\bibinfo {author} {\bibfnamefont {J.}~\bibnamefont
  {Paulose}}, \bibinfo {author} {\bibfnamefont {A.~S.}\ \bibnamefont
  {Meeussen}}, \ and\ \bibinfo {author} {\bibfnamefont {V.}~\bibnamefont
  {Vitelli}},\ }\href@noop {} {\bibfield  {journal} {\bibinfo  {journal} {Proc.
  Natl. Acad. Sci. U.S.A.}\ }\textbf {\bibinfo {volume} {112}},\ \bibinfo
  {pages} {7639} (\bibinfo {year} {2015})}\BibitemShut {NoStop}%
\bibitem [{\citenamefont {Carroll}\ \emph {et~al.}(2003)\citenamefont
  {Carroll}, \citenamefont {Levasseur}, \citenamefont {Wood}, \citenamefont
  {Whitaker}, \citenamefont {Jones},\ and\ \citenamefont
  {McDougall}}]{carroll2003exploring}%
  \BibitemOpen
  \bibfield  {author} {\bibinfo {author} {\bibfnamefont {M.}~\bibnamefont
  {Carroll}}, \bibinfo {author} {\bibfnamefont {M.}~\bibnamefont {Levasseur}},
  \bibinfo {author} {\bibfnamefont {C.}~\bibnamefont {Wood}}, \bibinfo {author}
  {\bibfnamefont {M.}~\bibnamefont {Whitaker}}, \bibinfo {author}
  {\bibfnamefont {K.~T.}\ \bibnamefont {Jones}}, \ and\ \bibinfo {author}
  {\bibfnamefont {A.}~\bibnamefont {McDougall}},\ }\href@noop {} {\bibfield
  {journal} {\bibinfo  {journal} {J. Cell Sci.}\ }\textbf {\bibinfo {volume}
  {116}},\ \bibinfo {pages} {4997} (\bibinfo {year} {2003})}\BibitemShut
  {NoStop}%
\bibitem [{\citenamefont {Bischof}\ \emph {et~al.}(2017)\citenamefont
  {Bischof}, \citenamefont {Brand}, \citenamefont {Somogyi}, \citenamefont
  {M\'ajer}, \citenamefont {Thome}, \citenamefont {Mori}, \citenamefont
  {Schwarz},\ and\ \citenamefont {L{\'e}n\'art}}]{Bischof_NCOMM}%
  \BibitemOpen
  \bibfield  {author} {\bibinfo {author} {\bibfnamefont {J.}~\bibnamefont
  {Bischof}}, \bibinfo {author} {\bibfnamefont {C.~A.}\ \bibnamefont {Brand}},
  \bibinfo {author} {\bibfnamefont {K.}~\bibnamefont {Somogyi}}, \bibinfo
  {author} {\bibfnamefont {I.}~\bibnamefont {M\'ajer}}, \bibinfo {author}
  {\bibfnamefont {S.}~\bibnamefont {Thome}}, \bibinfo {author} {\bibfnamefont
  {M.}~\bibnamefont {Mori}}, \bibinfo {author} {\bibfnamefont {U.~S.}\
  \bibnamefont {Schwarz}}, \ and\ \bibinfo {author} {\bibfnamefont
  {P.}~\bibnamefont {L{\'e}n\'art}},\ }\href@noop {} {\bibfield  {journal}
  {\bibinfo  {journal} {Nat. Commun.}\ }\textbf {\bibinfo {volume} {8}},\
  \bibinfo {pages} {849} (\bibinfo {year} {2017})}\BibitemShut {NoStop}%
\bibitem [{\citenamefont {Bischof}(2016)}]{bischof2016molecular}%
  \BibitemOpen
  \bibfield  {author} {\bibinfo {author} {\bibfnamefont {J.}~\bibnamefont
  {Bischof}},\ }\emph {\bibinfo {title} {The molecular mechanism of surface
  contraction waves in the starfish oocyte}},\ \href@noop {} {Ph.D. thesis},\
  \bibinfo  {school} {Universit\"at Heidelberg} (\bibinfo {year}
  {2016})\BibitemShut {NoStop}%
\bibitem [{\citenamefont {Ciarlet}(2005)}]{Ciarlet}%
  \BibitemOpen
  \bibfield  {author} {\bibinfo {author} {\bibfnamefont {P.~G.}\ \bibnamefont
  {Ciarlet}},\ }\href@noop {} {\emph {\bibinfo {title} {An introduction to
  differential geometry with applications to elasticity}}}\ (\bibinfo
  {publisher} {Springer, Netherlands},\ \bibinfo {year} {2005})\BibitemShut
  {NoStop}%
\bibitem [{\citenamefont {Pezzulla}\ \emph {et~al.}(2017)\citenamefont
  {Pezzulla}, \citenamefont {Stoop}, \citenamefont {Jiang},\ and\ \citenamefont
  {Holmes}}]{pezzulla2017curvature}%
  \BibitemOpen
  \bibfield  {author} {\bibinfo {author} {\bibfnamefont {M.}~\bibnamefont
  {Pezzulla}}, \bibinfo {author} {\bibfnamefont {N.}~\bibnamefont {Stoop}},
  \bibinfo {author} {\bibfnamefont {X.}~\bibnamefont {Jiang}}, \ and\ \bibinfo
  {author} {\bibfnamefont {D.~P.}\ \bibnamefont {Holmes}},\ }\href {\doibase
  10.1098/rspa.2017.0087} {\bibfield  {journal} {\bibinfo  {journal} {Proc. R.
  Soc. A}\ }\textbf {\bibinfo {volume} {473}} (\bibinfo {year} {2017}),\
  10.1098/rspa.2017.0087}\BibitemShut {NoStop}%
\bibitem [{\citenamefont {Heer}\ \emph {et~al.}(2017)\citenamefont {Heer},
  \citenamefont {Miller}, \citenamefont {Chanet}, \citenamefont {Stoop},
  \citenamefont {Dunkel},\ and\ \citenamefont {Martin}}]{heer2017actomyosin}%
  \BibitemOpen
  \bibfield  {author} {\bibinfo {author} {\bibfnamefont {N.~C.}\ \bibnamefont
  {Heer}}, \bibinfo {author} {\bibfnamefont {P.~W.}\ \bibnamefont {Miller}},
  \bibinfo {author} {\bibfnamefont {S.}~\bibnamefont {Chanet}}, \bibinfo
  {author} {\bibfnamefont {N.}~\bibnamefont {Stoop}}, \bibinfo {author}
  {\bibfnamefont {J.}~\bibnamefont {Dunkel}}, \ and\ \bibinfo {author}
  {\bibfnamefont {A.~C.}\ \bibnamefont {Martin}},\ }\href@noop {} {\bibfield
  {journal} {\bibinfo  {journal} {Development}\ }\textbf {\bibinfo {volume}
  {144}},\ \bibinfo {pages} {146761} (\bibinfo {year} {2017})}\BibitemShut
  {NoStop}%
\bibitem [{\citenamefont {Klein}\ \emph {et~al.}(2007)\citenamefont {Klein},
  \citenamefont {Efrati},\ and\ \citenamefont {Sharon}}]{klein2007shaping}%
  \BibitemOpen
  \bibfield  {author} {\bibinfo {author} {\bibfnamefont {Y.}~\bibnamefont
  {Klein}}, \bibinfo {author} {\bibfnamefont {E.}~\bibnamefont {Efrati}}, \
  and\ \bibinfo {author} {\bibfnamefont {E.}~\bibnamefont {Sharon}},\
  }\href@noop {} {\bibfield  {journal} {\bibinfo  {journal} {Science}\ }\textbf
  {\bibinfo {volume} {315}},\ \bibinfo {pages} {1116} (\bibinfo {year}
  {2007})}\BibitemShut {NoStop}%
\bibitem [{\citenamefont {Efrati}\ \emph {et~al.}(2009)\citenamefont {Efrati},
  \citenamefont {Sharon},\ and\ \citenamefont {Kupferman}}]{Efrati2009}%
  \BibitemOpen
  \bibfield  {author} {\bibinfo {author} {\bibfnamefont {E.}~\bibnamefont
  {Efrati}}, \bibinfo {author} {\bibfnamefont {E.}~\bibnamefont {Sharon}}, \
  and\ \bibinfo {author} {\bibfnamefont {R.}~\bibnamefont {Kupferman}},\ }\href
  {\doibase http://dx.doi.org/10.1016/j.jmps.2008.12.004} {\bibfield  {journal}
  {\bibinfo  {journal} {J. Mech. Phys. Solids}\ }\textbf {\bibinfo {volume}
  {57}},\ \bibinfo {pages} {762 } (\bibinfo {year} {2009})}\BibitemShut
  {NoStop}%
\bibitem [{\citenamefont {Kac}(1974)}]{kac1974stochastic}%
  \BibitemOpen
  \bibfield  {author} {\bibinfo {author} {\bibfnamefont {M.}~\bibnamefont
  {Kac}},\ }\href@noop {} {\bibfield  {journal} {\bibinfo  {journal} {Rocky
  Mountain J. Math.}\ }\textbf {\bibinfo {volume} {4}},\ \bibinfo {pages} {497}
  (\bibinfo {year} {1974})}\BibitemShut {NoStop}%
\bibitem [{\citenamefont {Thomson}(1854)}]{thomson1854theory}%
  \BibitemOpen
  \bibfield  {author} {\bibinfo {author} {\bibfnamefont {W.}~\bibnamefont
  {Thomson}},\ }\href@noop {} {\bibfield  {journal} {\bibinfo  {journal}
  {‎Proc. R. Soc. A}\ }\textbf {\bibinfo {volume} {7}},\ \bibinfo {pages}
  {382} (\bibinfo {year} {1854})}\BibitemShut {NoStop}%
\bibitem [{\citenamefont {Masoliver}\ and\ \citenamefont
  {Weiss}(1996)}]{masoliver1996finite}%
  \BibitemOpen
  \bibfield  {author} {\bibinfo {author} {\bibfnamefont {J.}~\bibnamefont
  {Masoliver}}\ and\ \bibinfo {author} {\bibfnamefont {G.~H.}\ \bibnamefont
  {Weiss}},\ }\href@noop {} {\bibfield  {journal} {\bibinfo  {journal} {Eur. J.
  Phys.}\ }\textbf {\bibinfo {volume} {17}},\ \bibinfo {pages} {190} (\bibinfo
  {year} {1996})}\BibitemShut {NoStop}%
\bibitem [{\citenamefont {Cirak}\ \emph {et~al.}(2000)\citenamefont {Cirak},
  \citenamefont {Ortiz},\ and\ \citenamefont {Schroder}}]{Cirak}%
  \BibitemOpen
  \bibfield  {author} {\bibinfo {author} {\bibfnamefont {F.}~\bibnamefont
  {Cirak}}, \bibinfo {author} {\bibfnamefont {M.}~\bibnamefont {Ortiz}}, \ and\
  \bibinfo {author} {\bibfnamefont {P.}~\bibnamefont {Schroder}},\ }\href@noop
  {} {\bibfield  {journal} {\bibinfo  {journal} {Int. J. Numer. Meth. Eng.}\
  }\textbf {\bibinfo {volume} {47}},\ \bibinfo {pages} {2039} (\bibinfo {year}
  {2000})}\BibitemShut {NoStop}%
\bibitem [{\citenamefont {Stoop}\ \emph {et~al.}(2015)\citenamefont {Stoop},
  \citenamefont {Lagrange}, \citenamefont {Terwagne}, \citenamefont {Reis},\
  and\ \citenamefont {Dunkel}}]{stoop2015curvature}%
  \BibitemOpen
  \bibfield  {author} {\bibinfo {author} {\bibfnamefont {N.}~\bibnamefont
  {Stoop}}, \bibinfo {author} {\bibfnamefont {R.}~\bibnamefont {Lagrange}},
  \bibinfo {author} {\bibfnamefont {D.}~\bibnamefont {Terwagne}}, \bibinfo
  {author} {\bibfnamefont {P.~M.}\ \bibnamefont {Reis}}, \ and\ \bibinfo
  {author} {\bibfnamefont {J.}~\bibnamefont {Dunkel}},\ }\href@noop {}
  {\bibfield  {journal} {\bibinfo  {journal} {Nat. Mater.}\ }\textbf {\bibinfo
  {volume} {14}},\ \bibinfo {pages} {337} (\bibinfo {year} {2015})}\BibitemShut
  {NoStop}%
\bibitem [{\citenamefont {Cheer}\ \emph {et~al.}(1987)\citenamefont {Cheer},
  \citenamefont {Vincent}, \citenamefont {Nuccitelli},\ and\ \citenamefont
  {Oster}}]{cheer1987cortical}%
  \BibitemOpen
  \bibfield  {author} {\bibinfo {author} {\bibfnamefont {A.}~\bibnamefont
  {Cheer}}, \bibinfo {author} {\bibfnamefont {J.-P.}\ \bibnamefont {Vincent}},
  \bibinfo {author} {\bibfnamefont {R.}~\bibnamefont {Nuccitelli}}, \ and\
  \bibinfo {author} {\bibfnamefont {G.}~\bibnamefont {Oster}},\ }\href@noop {}
  {\bibfield  {journal} {\bibinfo  {journal} {J. Theor. Biol.}\ }\textbf
  {\bibinfo {volume} {124}},\ \bibinfo {pages} {377} (\bibinfo {year}
  {1987})}\BibitemShut {NoStop}%
\bibitem [{\citenamefont {Taniguchi}\ \emph {et~al.}(2013)\citenamefont
  {Taniguchi}, \citenamefont {Ishihara}, \citenamefont {Oonuki}, \citenamefont
  {Honda-Kitahara}, \citenamefont {Kaneko},\ and\ \citenamefont
  {Sawai}}]{taniguchi2013phase}%
  \BibitemOpen
  \bibfield  {author} {\bibinfo {author} {\bibfnamefont {D.}~\bibnamefont
  {Taniguchi}}, \bibinfo {author} {\bibfnamefont {S.}~\bibnamefont {Ishihara}},
  \bibinfo {author} {\bibfnamefont {T.}~\bibnamefont {Oonuki}}, \bibinfo
  {author} {\bibfnamefont {M.}~\bibnamefont {Honda-Kitahara}}, \bibinfo
  {author} {\bibfnamefont {K.}~\bibnamefont {Kaneko}}, \ and\ \bibinfo {author}
  {\bibfnamefont {S.}~\bibnamefont {Sawai}},\ }\href@noop {} {\bibfield
  {journal} {\bibinfo  {journal} {Proc. Natl. Acad. Sci. U.S.A.}\ }\textbf
  {\bibinfo {volume} {110}},\ \bibinfo {pages} {5016} (\bibinfo {year}
  {2013})}\BibitemShut {NoStop}%
\bibitem [{\citenamefont {Bement}\ \emph {et~al.}(2015)\citenamefont {Bement},
  \citenamefont {Leda}, \citenamefont {Moe}, \citenamefont {Kita},
  \citenamefont {Larson}, \citenamefont {Golding}, \citenamefont {Pfeuti},
  \citenamefont {Su}, \citenamefont {Miller}, \citenamefont {Goryachev} \emph
  {et~al.}}]{bement2015activator}%
  \BibitemOpen
  \bibfield  {author} {\bibinfo {author} {\bibfnamefont {W.~M.}\ \bibnamefont
  {Bement}}, \bibinfo {author} {\bibfnamefont {M.}~\bibnamefont {Leda}},
  \bibinfo {author} {\bibfnamefont {A.~M.}\ \bibnamefont {Moe}}, \bibinfo
  {author} {\bibfnamefont {A.~M.}\ \bibnamefont {Kita}}, \bibinfo {author}
  {\bibfnamefont {M.~E.}\ \bibnamefont {Larson}}, \bibinfo {author}
  {\bibfnamefont {A.~E.}\ \bibnamefont {Golding}}, \bibinfo {author}
  {\bibfnamefont {C.}~\bibnamefont {Pfeuti}}, \bibinfo {author} {\bibfnamefont
  {K.-C.}\ \bibnamefont {Su}}, \bibinfo {author} {\bibfnamefont {A.~L.}\
  \bibnamefont {Miller}}, \bibinfo {author} {\bibfnamefont {A.~B.}\
  \bibnamefont {Goryachev}},  \emph {et~al.},\ }\href@noop {} {\bibfield
  {journal} {\bibinfo  {journal} {Nat. Cell Bio.}\ }\textbf {\bibinfo {volume}
  {17}},\ \bibinfo {pages} {1471} (\bibinfo {year} {2015})}\BibitemShut
  {NoStop}%
\bibitem [{\citenamefont {Goriely}\ \emph {et~al.}(2008)\citenamefont
  {Goriely}, \citenamefont {Robertson-Tessi}, \citenamefont {Tabor},\ and\
  \citenamefont {Vandiver}}]{goriely2008elastic}%
  \BibitemOpen
  \bibfield  {author} {\bibinfo {author} {\bibfnamefont {A.}~\bibnamefont
  {Goriely}}, \bibinfo {author} {\bibfnamefont {M.}~\bibnamefont
  {Robertson-Tessi}}, \bibinfo {author} {\bibfnamefont {M.}~\bibnamefont
  {Tabor}}, \ and\ \bibinfo {author} {\bibfnamefont {R.}~\bibnamefont
  {Vandiver}},\ }in\ \href@noop {} {\emph {\bibinfo {booktitle} {Mathematical
  modelling of biosystems}}}\ (\bibinfo  {publisher} {Springer},\ \bibinfo
  {year} {2008})\ pp.\ \bibinfo {pages} {1--44}\BibitemShut {NoStop}%
\bibitem [{\citenamefont {Polthier}\ and\ \citenamefont
  {Schmies}(1999)}]{polthier1999geodesic}%
  \BibitemOpen
  \bibfield  {author} {\bibinfo {author} {\bibfnamefont {K.}~\bibnamefont
  {Polthier}}\ and\ \bibinfo {author} {\bibfnamefont {M.}~\bibnamefont
  {Schmies}},\ }\enquote {\bibinfo {title} {Geodesic flow on polyhedral
  surfaces},}\ in\ \href@noop {} {\emph {\bibinfo {booktitle} {Data
  Visualization '99: Proceedings of the Joint EUROGRAPHICS and IEEE TCVG
  Symposium on Visualization in Vienna, Austria, May 26--28, 1999}}},\ \bibinfo
  {editor} {edited by\ \bibinfo {editor} {\bibfnamefont {E.}~\bibnamefont
  {Gr{\"o}ller}}, \bibinfo {editor} {\bibfnamefont {H.}~\bibnamefont
  {L{\"o}ffelmann}}, \ and\ \bibinfo {editor} {\bibfnamefont {W.}~\bibnamefont
  {Ribarsky}}}\ (\bibinfo  {publisher} {Springer, {V}ienna},\ \bibinfo {year}
  {1999})\ pp.\ \bibinfo {pages} {179--188}\BibitemShut {NoStop}%
\bibitem [{SI()}]{SI}%
  \BibitemOpen
  \href@noop {} {}\bibinfo {note} {See Supplementary Material, which includes
  Refs. \cite{goriely2005differential}, \cite{quilliet2008anisotropic}, and
  \cite{vaziri2008localized}}\BibitemShut {NoStop}%
\bibitem [{\citenamefont {Jones}(1984)}]{jones1984stability}%
  \BibitemOpen
  \bibfield  {author} {\bibinfo {author} {\bibfnamefont {C.~K. R.~T.}\
  \bibnamefont {Jones}},\ }\href@noop {} {\bibfield  {journal} {\bibinfo
  {journal} {Trans. Amer. Math. Soc}\ }\textbf {\bibinfo {volume} {286}},\
  \bibinfo {pages} {431} (\bibinfo {year} {1984})}\BibitemShut {NoStop}%
\bibitem [{\citenamefont {Shah}\ and\ \citenamefont
  {Keren}(2014)}]{shah2014symmetry}%
  \BibitemOpen
  \bibfield  {author} {\bibinfo {author} {\bibfnamefont {E.~A.}\ \bibnamefont
  {Shah}}\ and\ \bibinfo {author} {\bibfnamefont {K.}~\bibnamefont {Keren}},\
  }\href@noop {} {\bibfield  {journal} {\bibinfo  {journal} {eLife}\ }\textbf
  {\bibinfo {volume} {3}},\ \bibinfo {pages} {e01433} (\bibinfo {year}
  {2014})}\BibitemShut {NoStop}%
\bibitem [{\citenamefont {{Tan}}\ \emph {et~al.}()\citenamefont {{Tan}},
  \citenamefont {{Malik Garbi}}, \citenamefont {{Abu-Shah}}, \citenamefont
  {{Li}}, \citenamefont {{Sharma}}, \citenamefont {{MacKintosh}}, \citenamefont
  {{Keren}}, \citenamefont {{Schmidt}},\ and\ \citenamefont
  {{Fakhri}}}]{2016arXiv160307600T}%
  \BibitemOpen
  \bibfield  {author} {\bibinfo {author} {\bibfnamefont {T.~H.}\ \bibnamefont
  {{Tan}}}, \bibinfo {author} {\bibfnamefont {M.}~\bibnamefont {{Malik
  Garbi}}}, \bibinfo {author} {\bibfnamefont {E.}~\bibnamefont {{Abu-Shah}}},
  \bibinfo {author} {\bibfnamefont {J.}~\bibnamefont {{Li}}}, \bibinfo {author}
  {\bibfnamefont {A.}~\bibnamefont {{Sharma}}}, \bibinfo {author}
  {\bibfnamefont {F.~C.}\ \bibnamefont {{MacKintosh}}}, \bibinfo {author}
  {\bibfnamefont {K.}~\bibnamefont {{Keren}}}, \bibinfo {author} {\bibfnamefont
  {C.~F.}\ \bibnamefont {{Schmidt}}}, \ and\ \bibinfo {author} {\bibfnamefont
  {N.}~\bibnamefont {{Fakhri}}},\ }\href@noop {} {\bibinfo  {journal}
  {arXiv:1603.07600}\ }\BibitemShut {NoStop}%
\bibitem [{\citenamefont {Yoshida}(2010)}]{yoshida2010self}%
  \BibitemOpen
\bibfield  {journal} {  }\bibfield  {author} {\bibinfo {author} {\bibfnamefont
  {R.}~\bibnamefont {Yoshida}},\ }\href@noop {} {\bibfield  {journal} {\bibinfo
   {journal} {Adv. Mater.}\ }\textbf {\bibinfo {volume} {22}},\ \bibinfo
  {pages} {3463} (\bibinfo {year} {2010})}\BibitemShut {NoStop}%
\bibitem [{\citenamefont {Goriely}\ and\ \citenamefont
  {Ben~Amar}(2005)}]{goriely2005differential}%
  \BibitemOpen
  \bibfield  {author} {\bibinfo {author} {\bibfnamefont {A.}~\bibnamefont
  {Goriely}}\ and\ \bibinfo {author} {\bibfnamefont {M.}~\bibnamefont
  {Ben~Amar}},\ }\href@noop {} {\bibfield  {journal} {\bibinfo  {journal}
  {Phys. Rev. Lett.}\ }\textbf {\bibinfo {volume} {94}},\ \bibinfo {pages}
  {198103} (\bibinfo {year} {2005})}\BibitemShut {NoStop}%
\bibitem [{\citenamefont {Quilliet}\ \emph {et~al.}(2008)\citenamefont
  {Quilliet}, \citenamefont {Zoldesi}, \citenamefont {Riera}, \citenamefont
  {Van~Blaaderen},\ and\ \citenamefont {Imhof}}]{quilliet2008anisotropic}%
  \BibitemOpen
  \bibfield  {author} {\bibinfo {author} {\bibfnamefont {C.}~\bibnamefont
  {Quilliet}}, \bibinfo {author} {\bibfnamefont {C.}~\bibnamefont {Zoldesi}},
  \bibinfo {author} {\bibfnamefont {C.}~\bibnamefont {Riera}}, \bibinfo
  {author} {\bibfnamefont {A.}~\bibnamefont {Van~Blaaderen}}, \ and\ \bibinfo
  {author} {\bibfnamefont {A.}~\bibnamefont {Imhof}},\ }\href@noop {}
  {\bibfield  {journal} {\bibinfo  {journal} {Eur. Phys. J. E}\ }\textbf
  {\bibinfo {volume} {27}},\ \bibinfo {pages} {13} (\bibinfo {year}
  {2008})}\BibitemShut {NoStop}%
\bibitem [{\citenamefont {Vaziri}\ and\ \citenamefont
  {Mahadevan}(2008)}]{vaziri2008localized}%
  \BibitemOpen
  \bibfield  {author} {\bibinfo {author} {\bibfnamefont {A.}~\bibnamefont
  {Vaziri}}\ and\ \bibinfo {author} {\bibfnamefont {L.}~\bibnamefont
  {Mahadevan}},\ }\href@noop {} {\bibfield  {journal} {\bibinfo  {journal}
  {Proc. Natl. Acad. Sci. U.S.A.}\ }\textbf {\bibinfo {volume} {105}},\
  \bibinfo {pages} {7913} (\bibinfo {year} {2008})}\BibitemShut {NoStop}%
\end{thebibliography}
%
	
%

\end{document}